\documentclass[final,5p,times,twocolumn]{elsarticle}
\usepackage{amsmath,amssymb,bm,graphicx,epsfig}
\RequirePackage[colorlinks,citecolor=blue,urlcolor=blue,linkcolor=blue]{hyperref}

\def\eps{\epsilon}
\def\Mgg{M_{\gamma \gamma}}

\newcommand\as{\alpha_S} 
\newcommand\muR{\mu_{\rm R}}
\newcommand\muF{\mu_{\rm F}}

\def\beqn{\begin{eqnarray}} 
\def\eeqn{\end{eqnarray}} 
\def\beq{\begin{equation}} 
\def\eeq{\end{equation}}

\biboptions{square,sort&compress} 

\journal{Physics Letters B}

\begin{document}

\begin{frontmatter}

\title{Full top-quark mass dependence in diphoton production at NNLO in QCD}
  
\author[a,a1]{Matteo Becchetti}

\address[a]{Dipartimento di Fisica, Università di Torino and INFN Sezione di Torino, Via Pietro Giuria 1, I-10125 Torino, Italy}

\ead[a1]{matteo.becchetti@unito.it}

\author[b,b1]{Roberto Bonciani}

\address[b]{Dipartimento di Fisica, Università di Roma “La Sapienza” and
INFN Sezione di Roma, Piazzale Aldo Moro 2, I-00185 Roma, Italy}

\ead[b1]{roberto.bonciani@roma1.infn.it}

\author[c,c1]{Leandro Cieri}

\ead[c1]{lcieri@ific.uv.es}

\author[c,c2]{Federico Coro}

\ead[c2]{fcoro@ific.uv.es}

\author[b,b2]{Federico Ripani}

\ead[b2]{federico.ripani@uniroma1.it}

\address[c]{Instituto de Física Corpuscular, Universitat de València - Consejo Superior de
Investigaciones Cientificas, Parc Cientific, E-46980 Paterna, Valencia, Spain}

\begin{abstract}
In this paper we consider the diphoton production in hadronic collisions at the next-to-next-to-leading order (NNLO) in perturbative QCD, taking into account for the first time the full top quark mass dependence up to two loops (full NNLO). We show selected numerical distributions, highlighting the kinematic regions where the massive corrections are more significant. We make use of the recently computed two-loop massive amplitudes for diphoton production in the quark annihilation channel. The remaining massive contributions at NNLO are also considered, and we comment on the weight of the different types of contributions to the full and complete result.    
\end{abstract}


\end{frontmatter}

\section{Introduction}
\label{sec:intro}

The production of two isolated prompt photons (diphotons) remains one of the most important processes at the Large Hadron Collider (LHC). It is a probe of the Standard Model (SM) of particle physics \cite{ATLAS:2021mbt,ATLAS:2017cvh,CMS:2014mvm,ATLAS:2012fgo} and was one of the two most important channels in searches and studies of the Higgs boson \cite{ATLAS:2012yve,CMS:2012qbp,ATLAS:2022fnp,CMS:2022wpo,CMS:2021kom,CMS:2020xrn,ATLAS:2018hxb,CMS:2014afl,ATLAS:2014cnc,ATLAS:2014yga}. 

Several new physics searches \cite{ATLAS:2023hbp,ATLAS:2023meo,ATLAS:2022abz,CMS:2019pov,CMS:2016kgr,ATLAS:2017ayi} are still being carried out using the diphoton channel, due to the clean signature of the photons in the LHC electromagnetic calorimeters. 

Owing to its phenomenological relevance, precise theoretical results are required in order to compare with the LHC data. The state of the art for diphoton production is the next-to-next-to-leading order (NNLO) accuracy (taking into account five light quark flavours) \cite{Catani:2011qz,Campbell:2016yrh,Catani:2018krb,Schuermann:2022qdm} in perturbative QCD. Although all the necessary elements of the (massless) next-to-next-to-next-to-leading order (N$^3$LO) are already known \cite{Bern:2001df,Caola:2022dfa,Chawdhry:2020for,Agarwal:2021grm,Chawdhry:2021mkw,Agarwal:2021vdh,Chawdhry:2021hkp,Badger:2021ohm}, they have not yet been included together in order to obtain phenomenological results at full N$^3$LO accuracy. First-order electroweak/QED corrections are also known \cite{Chiesa:2017gqx,Cieri:2021fdb}.

The calculations at NLO include fragmentation contributions at the same level of accuracy~\cite{Binoth:1999qq} and transverse momentum resummation at the corresponding precision, the next-to-leading order logarithmic accuracy (NLL) \cite{Balazs:2007hr}. At that time, the so-called box (one-loop $gg\rightarrow\gamma\gamma$) contribution was already known \cite{Dicus:1987fk}.
Due to the large gluon luminosity at the LHC, the size of the box contribution is of the order of the Born sub-process ($q\bar{q} \rightarrow \gamma \gamma$) and for this reason, although formally of order $\mathcal{O}(\as^2)$, it was considered in the NLO analyses~
\cite{Bern:2002jx,Campbell:2011bn}. 

Regarding the diphoton background in Higgs boson production, it is possible to constrain the Higgs boson width from interference effects of the continuum $gg\to \gamma\gamma$ spectrum with the signal $gg\to H\to \gamma\gamma$. The phenomenology behind this process has therefore been studied in detail in the literature, with effective calculations at NLO (and beyond)~\cite{Dicus:1987fk,Dixon:2003yb,Martin:2012xc,deFlorian:2013psa,Martin:2013ula,Dixon:2013haa,Campbell:2017rke,Cieri:2017kpq,Bargiela:2022dla}.

The small transverse momentum region of the diphoton pair is also of interest in SM and Higgs boson studies, in the determination of the Higgs boson width, etc. The transverse momentum ($q_T$) resummation for diphoton production is known at next-to-next-to-leading logarithmic accuracy (NNLL) \cite{Cieri:2015rqa} and at N$^3$LL \cite{Becher:2020ugp} in association with fixed-order NNLO results.

The possibility of measuring the top quark mass has been pointed out in the literature~\cite{Jain:2016kai,Kawabata:2016aya}, if massive scattering amplitudes in diphoton production are taken into account (via loop corrections). These threshold effects of top quark pair production are manifested in the diphoton-invariant mass spectrum around two times the value of the top quark mass.

Non-trivial QCD corrections, including the dependence on the top quark mass, appear for the first time at NNLO. In ref. \cite{Campbell:2016yrh}, the one-loop $gg\rightarrow\gamma\gamma$ scattering amplitude was considered taking into account the top quark mass dependence (together with partial N$^3$LO contributions).

Regarding the inclusion of corrections beyond the NNLO ($\mathcal{O}(\as^3)$), the simplest approach for the gluon fusion channel is to consider the NLO effective QCD corrections to the box contribution (which captures some of the largest contributions). These NLO corrections form a gauge invariant subset \cite{Bern:2002jx} of the whole N$^3$LO gluon fusion channel. In the massless case, the NLO corrections to the box contribution have been calculated in ref. \cite{Bern:2002jx} using the massless two-loop scattering amplitudes of ref. \cite{Bern:2001df}. In the context of the massive contributions, two recent papers have shown the impact of these massive corrections on the gluon fusion channel \cite{Maltoni:2018zvp,Chen:2019fla}, which turned out to be sizable.

Even in the light of previous efforts to calculate the scattering amplitudes that capture the most significant contributions to diphoton production, the full massive QCD NNLO corrections for this process are still missing (mainly due to the previously unknown two-loop scattering amplitudes \cite{Becchetti:2023wev}).

In this paper, we consider for the first time diphoton production at NNLO, taking into account the full top quark mass dependence. We include all NNLO massive scattering amplitudes: i) the box contribution in the gluon fusion channel~\cite{Campbell:2016yrh}, ii) the two-loop scattering amplitudes\footnote{The two-loop amplitude is UV renormalized in five-flavors decoupling scheme.} $q\bar{q} \rightarrow \gamma \gamma$ \cite{Becchetti:2023wev}, and the real radiation contributions (double real and real-virtual).

The paper is organised as follows. In Section 2 we explain the setup of our calculation. In Section 3 we present selected numerical results for the LHC phenomenology, and in Section 4 we present our conclusions.

\section{Organisation of the calculation}
\label{sec:calculation}

Since the massless (five light quark flavours) NNLO QCD corrections to diphoton production are known \cite{Catani:2011qz}, our approach is to consider all the remaining massive scattering amplitudes and combine them in an appropriate way.

\begin{figure}
\centering
  \includegraphics[width=0.5\textwidth, angle=360]{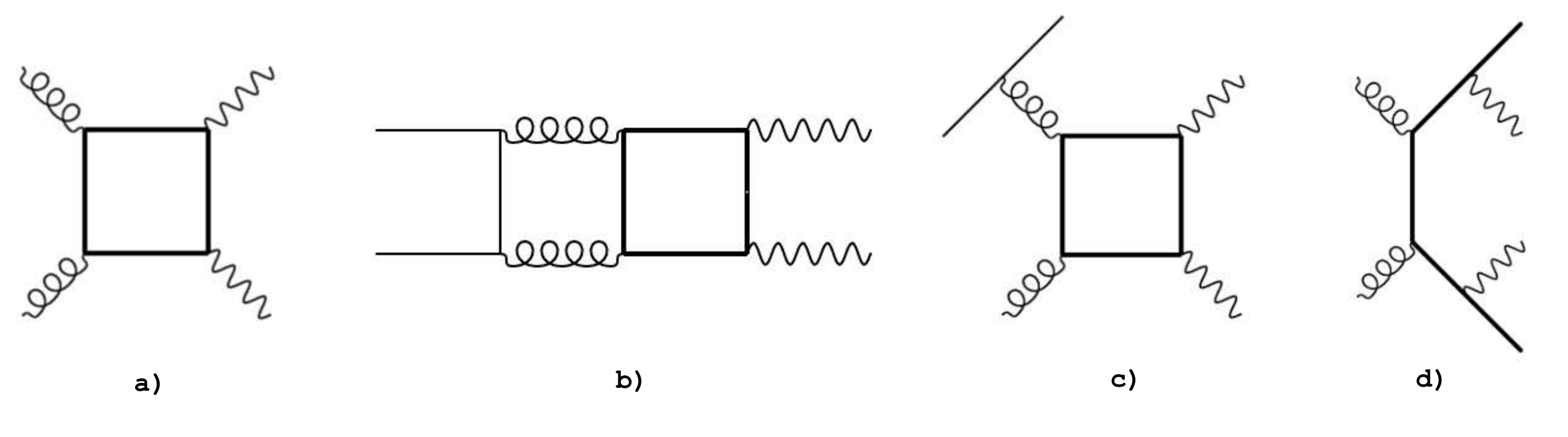}
\caption
{ \label{fig:contrib}
  { \em Different types of contributions to the massive corrections at NNLO for diphoton production in perturbative QCD. The explanation of the different features is given in the text.}}
\end{figure}

The first non-trivial massive corrections appear at NNLO. We classify the scattering amplitudes into four types of contributions. In first place, we consider the known massive one-loop box scattering amplitude $gg\rightarrow\gamma\gamma$ \cite{Campbell:2016yrh} as depicted in Fig. \ref{fig:contrib} \texttt{a)}. The two-loop (double-virtual) corrections to the Born sub-process $q\bar{q} \rightarrow \gamma \gamma$ \cite{Becchetti:2023wev} are shown with a representative Feynman diagram in Fig. \ref{fig:contrib} \texttt{b)}, where in the loop we consider a massive top quark. We also consider massive real-virtual contributions to diphoton production (see Fig. \ref{fig:contrib} \texttt{c)}), where the diphoton pair is produced in association with real radiation (quarks and gluons). This scattering amplitude is interfered with the corresponding tree-level matrix element ($q\bar{q} \rightarrow \gamma \gamma g $ or $qg \rightarrow \gamma \gamma q $ depending on the partonic channel). The partonic contributions to Fig. \ref{fig:contrib} \texttt{c)} are finite, not only in four dimensions, but also after integration over the transverse momentum of the diphoton pair ($p_{T}^{\gamma\gamma}$). This amplitude is presented in the appendix of ref. \cite{Campbell:2016yrh}, but it is considered for its squared modulus (effective N$^3$LO contribution). In our case we calculated this contribution and we checked it numerically with \texttt{OpenLoops} \cite{Buccioni:2019sur,Cascioli:2011va,Zoller:2018zwf,vanHameren:2009dr,vanHameren:2010cp,Denner:2016kdg}. The last element that we considered (which completes the NNLO massive corrections) is shown in Fig. \ref{fig:contrib} \texttt{d)}; it is related to diphoton production in association to the emission of two on-shell top-quarks ($q\bar{q} \rightarrow \gamma\gamma t \bar{t}$ and $gg \rightarrow \gamma\gamma t \bar{t}$). We computed these double-real amplitudes and we checked them numerically with \texttt{OpenLoops}. Although this sub-process can be effectively detected experimentally and (therefore) then subtracted, we include it explicitly in our calculation. Indeed, LHC measurements of diphoton production take into account any kind of additional radiation accompanying the two isolated photons, and therefore this contribution must be taken into account in any comparison with LHC data \cite{ATLAS:2021mbt,ATLAS:2017cvh,CMS:2014mvm,ATLAS:2012fgo} that claim full NNLO QCD massive corrections.

All our massive corrections are encoded in a new version of the \texttt{$2\gamma$NNLO} code \cite{Catani:2011qz}, which has been cross-checked with the \texttt{MATRIX} \cite{Grazzini:2017mhc} numerical code (version 2.0.0) (which includes the massless NNLO QCD corrections to diphoton production). The new version of the \texttt{$2\gamma$NNLO} code benefits from the fast integration routines of the \texttt{DYTurbo} framework \cite{Camarda:2019zyx,Camarda:2021ict}.

The double-real and real-virtual sub-processes (see Fig. \ref{fig:contrib} \texttt{c)} and \texttt{d)}) are not only finite in four dimensions, but they are also finite after integration over the transverse momentum of the diphoton pair. We have checked numerically that under $q_T$ integration these contributions are finite and numerically stable in the whole $q_T$ range.

\section{NNLO results with full top-quark mass dependence}

In this section, we present our results for the diphoton production at NNLO in perturbative QCD, taking into account the full top-quark mass dependence. We fix the pole mass $m_t$ of the top quark to the value $m_t = 173$~GeV. Our computational setup that was explained in Sec. \ref{sec:calculation}, has been encoded in a new version of the \texttt{$2\gamma$NNLO} code.

We consider isolated diphoton production in $pp$ collisions at the centre--of--mass energy $\sqrt{ s}=13$~TeV. We apply the following kinematical cuts on photon transverse momenta and rapidities: $p_{T \gamma}^{\rm hard} \geq 40$~GeV, $p_{T \gamma}^{\rm soft}\geq 30$~GeV and the rapidity of both photons is limited in the range $|y_\gamma|<2.37$, excluding the rapidity interval $1.37 < |y_\gamma|< 1.52$. The minimum angular separation between the two photons is $R_{\gamma \gamma}^{\rm min}=0.4$. These are essentially the kinematical cuts used in the ATLAS Collaboration study of ref.~\cite{ATLAS:2021mbt}. 

In the perturbative calculation, the QED coupling constant $\alpha$ is fixed at $1/\alpha = 137.035999139$. We use the central set of the \verb|NNPDF3.1| PDFs~\cite{NNPDF:2017mvq} as implemented in the \texttt{LHAPDF} framework~\cite{Buckley:2014ana} and the associated strong coupling with $\as(M_Z)=0.118$.

The central factorization and renormalization scale is chosen to be equal to the invariant mass of the diphoton pair $\mu \equiv \muR = \muF =  M_{\gamma\gamma}$. The theoretical uncertainty is estimated by varying the default scale choice for $\muR$ and $\muF$ independently by factors of $\{1/2,2\}$, while omitting combinations with $\muR/\muF = 4$ or $1/4$, resulting in the usual seven-point variation of scale combinations. Our standard choice of the central scale, can be replaced with other options, for instance the transverse mass of the diphoton pair, $M^{\rm T}_{\gamma\gamma} = \sqrt{(\Mgg)^2+(p_{T}^{\gamma\gamma})^2}$. Since our aim is to show the impact of the new massive corrections, we refer the reader to more detailed studies on scale variation (and scale choices) to refs. \cite{Catani:2018krb,Schuermann:2022qdm}.

We use the smooth cone isolation criterion \cite{Frixione:1998jh} (see also refs.~\cite{Frixione:1999gr,Catani:2000jh,Catani:2018krb}), which fixes the size $R$ of the isolation cone (drawn around the direction of the photon) and requires that the hadronic activity $E_{T}^{had}$ allowed inside the cone satisfies 
\begin{eqnarray}
\label{Eq:Isol_frixcriterion} 
 E_{T}^{had}(r) \leq \,\eps~p_{T \gamma}~\chi(r;R) \;, \quad
 \mbox{ in {\it all} cones with} \;r \leq R 
\;\;,    
\end{eqnarray}
where the function $\chi(r;R)$ is defined as  
\begin{equation}
\label{Eq:Isol_chirpow}
\chi(r;R) = \left( \frac{r}{R} \right)^{2n}\;.
\end{equation}
The specific values of the isolation parameters in our case~\cite{ATLAS:2021mbt} are: $R=0.4$, $\eps = 0.09$ and we take $n=1$. The choice of the $\chi(r;R)$ function as well as the particular value of the exponent $n$ is explained in ref. \cite{Catani:2018krb}. Since our aim is to present the effects of massive corrections, we suggest that the interested reader consult the isolation studies in refs. \cite{Catani:2018krb,Schuermann:2022qdm}. The top quark threshold region ($\sim 346$ GeV) is not particularly sensitive to the effects of the choice of isolation parameters \cite{Catani:2018krb}.

\begin{figure}
\centering
  \includegraphics[width=0.5\textwidth]{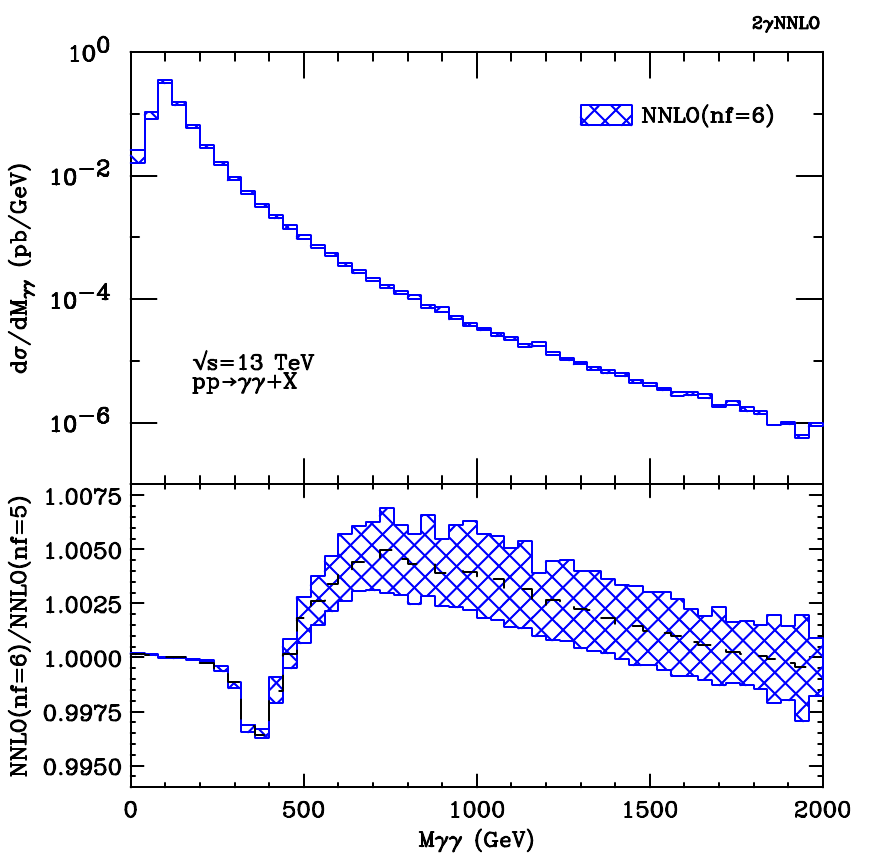}
\caption
{ \label{fig:invmassnnlo}
  { \em NNLO invariant mass distribution with full top quark mass dependence. In the lower panel we plot the ratio of the NNLO invariant mass distribution between the massive result and that with only five light quark flavours. The bands are obtained (as explained in the text) using the customary 7-point scale variation. The central scale is shown with a black dashed line.}}
\end{figure}

Since we rely on the $q_T$ subtraction method \cite{Catani:2007vq,Catani:2013tia} to perform our NNLO calculations, we use the technical parameter\footnote{We have also checked that in the extrapolation to $r_{\rm cut} =0$ the associated results for the total cross sections and the differential distributions vary less than $1\%$ when $r_{\rm cut} = 0.0005$ is used.} (a cut on the transverse momentum of the diphoton pair) $r_{\rm cut} = 0.0005 < p_{T}^{\gamma \gamma} / M_{\gamma\gamma}$. The large diphoton invariant mass tail is not particularly sensitive to $r_{\rm cut}$ variations around our chosen value \cite{Grazzini:2017mhc}. Studies on the impact of the fiducial power corrections and on the size of the $r_{\rm cut}$ parameter in colour singlet processes can be found in refs. \cite{Grazzini:2017mhc,Ebert:2019zkb,Camarda:2021jsw,Buonocore:2021tke}. 

The rest of this section proceeds as follows: first, we anticipate our results for diphoton production at NNLO in perturbative QCD, taking into account the full top quark mass dependence. At the end of this section, we discuss the relative weight of the different massive contributions involved in the NNLO calculation.

\begin{figure}
\centering
  \includegraphics[width=0.5\textwidth]{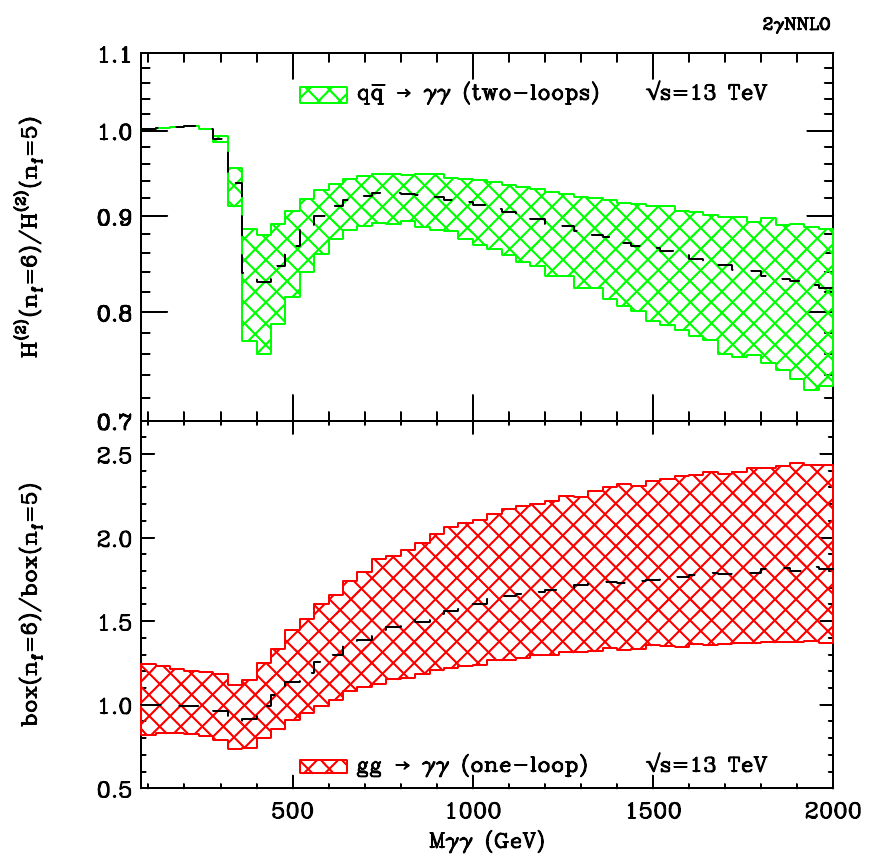}
\caption
{ \label{fig:H2andBox}
  { \em Ratios of diverse massive corrections to the massless case. In the upper panel we show the ratio of the two-loop ($q\bar{q} \rightarrow \gamma\gamma$) massive form factor to the massless one. In the lower panel we show the corresponding ratio but for the 1-loop box ($gg\rightarrow \gamma\gamma$) contribution. The central scale is shown with a black dashed line.}}
\end{figure}

\begin{figure}
\centering
  \includegraphics[width=0.5\textwidth]{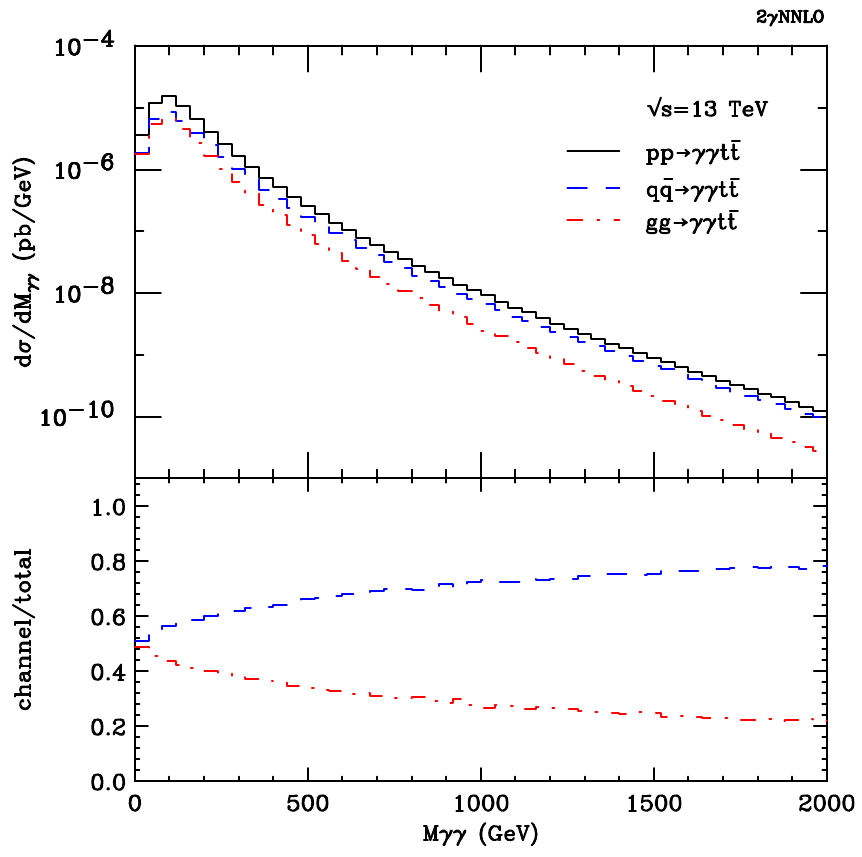}
\caption
{ \label{fig:RR}
  { \em Invariant mass distribution of the double-real ($pp \rightarrow \gamma\gamma t \bar{t}$) contribution to the NNLO fully massive result. In the lower panel we show the relative size of each one of the partonic channels that form the total double-real contribution. Only central scale results are shown.}}
\end{figure}

In Fig.~\ref{fig:invmassnnlo} we present our results regarding the invariant mass distribution of the photon pair at NNLO using the kinematical cuts described at the beginning of this section. In the lower panel, we show the ratio between the fully massive NNLO result and the NNLO prediction for five light quark flavours (5lf). Around the region $\Mgg \sim 2m_t$ (the top-quark pair threshold), the invariant mass distribution exhibits its negative peak \footnote{In the ratios, the corrections larger (smaller) than the unity are named \textit{positive} (\textit{negative}) since they are larger (smaller) than the five light flavour result.} due to a superposition of effects coming from the loop contributions. In the low-mass region ($\Mgg < 2m_t$), the massive result is still slightly larger than the massless case since the real corrections can resolve the top quark loop because the total centre--of--mass energy can be larger than $2m_t$ \cite{Maltoni:2018zvp}. Beyond the negative peak, the massive NNLO prediction presents its maximum (positive) deviation from the massless result at about 2.3 times of the value of the top quark pair threshold. The position (and shape) of this positive peak is the result of a competition between two opposite behaviours of (mainly) two contributions: the box scattering amplitude ($gg$-channel) and the two-loop form factor ($q\bar{q}$-channel) (see Fig. \ref{fig:H2andBox}). We postpone the discussion of the decreasing tail in the ratio between the massive and massless result, to the end of this section. The effect of the massive corrections (within the fiducial cuts discussed above) in the invariant mass from 1~GeV to 2~TeV is a deviation from the massless result in the range [-0.4\%, 0.8\%]. The effect may be larger if we use different selection cuts and for values of $M_{\gamma\gamma} > 2$ TeV.

We now comment on the contribution of the two-loop form factor to the NNLO invariant mass distribution\footnote{The representative Feynman diagram of this two-loop contribution is shown in Fig. \ref{fig:contrib} \texttt{b)}.}. In Fig.~\ref{fig:H2andBox} (upper panel) we show the ratio between the fully massive two-loop form factor and the massless case. The ratio is performed explicitly using the hard virtual factors $H^{(2)}$ defined in the \textit{hard resummation scheme} as explained in ref.~\cite{Catani:2013tia} and in our upcoming paper of the two-loop massive form factors \cite{Becchetti:2023wev}. The bands are computed implementing the usual 7-point scale variation as described at the beginning of this section. As in any massive loop contribution, the ratio exhibits the typical peak around the top quark threshold. For invariant masses larger than $2m_t$ (and after a peak around $\Mgg \sim 2.3 \times 2m_t$), the tail decreases (in part) since the two-loop massive corrections \footnote{More clearly, the separate contribution of the massive scattering amplitudes is negative (without considering the additional massless part).} are negative from invariant masses $\Mgg\sim 2 m_t$. At this point we observed also, that the massless two-loop form factor obtained with six light quark flavours is smaller (in the whole invariant mass range) than the result with five flavours. Moreover, the asymptotic behaviour (at large invariant masses) of the ratio between the two previous massless results ($H^{(2)} (n_f=6 {\rm lf})/H^{(2)} (n_f=5$)) is decreasing, as the corresponding behaviour with the massive result shown in Fig.~\ref{fig:H2andBox} top panel.

\begin{figure}
\centering
  \includegraphics[width=0.5\textwidth]{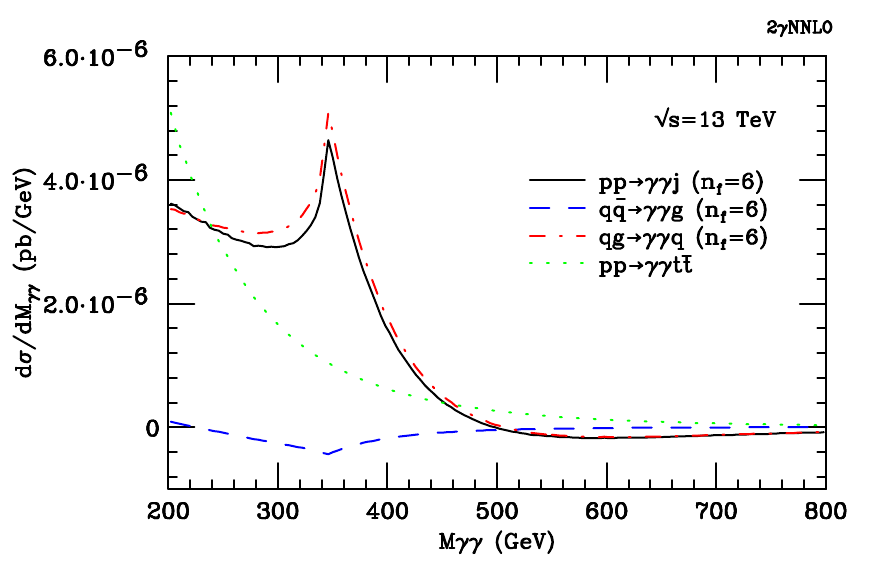}
\caption
{ \label{fig:RV}
  { \em Invariant mass distribution of the one-loop real emission massive contribution at NNLO. Only the massive top quark circulates in the loop. The light quark flavours are already considered in the massless part of the calculation. We show the different partonic channels and a comparison with the size of the double real correction shown in Fig. \ref{fig:RR}.}}
\end{figure}

In the bottom panel of Fig.~\ref{fig:H2andBox} we show the known behaviour of the ratio between the fully massive one-loop box contribution and the corresponding contribution with five light quark flavours. In this case, for large values of $\Mgg \gg m_t$, the ratio asymptotically\footnote{For the kinematical cuts considered here, the asymptotic regime is reached at (roughly) 3 TeV, which is not shown in these plots.} approaches the value \mbox{$(\sum_{nf=6} e_q^2)^2/(\sum_{n_f = 5} e_q^2)^2 = 225/121$}, implying that the massive contribution behaves as if it were composed of 6 light quark flavours \cite{Campbell:2016yrh}. The size of both ratios around the negative peak is quantitatively similar and amounts to roughly $-15\%$. These are the two most sizable massive contributions at NNLO accuracy. The distinctive and opposite behaviour of these two contributions at large values of $\Mgg$ (taking into account also the vanishing luminosity of the gluon) determines the position of the positive peak in the ratio of Fig.~\ref{fig:invmassnnlo}.

We now consider the massive double real corrections ($pp \rightarrow \gamma\gamma t \bar{t}$). In Fig.~\ref{fig:RR} we show the invariant mass distribution obtained from this partonic sub-process. Since we produce two on-shell top quarks, and since we are dealing with tree-level scattering amplitudes, there is no top quark threshold in the distribution (it has a continuously decreasing (logarithmic) tail as in the massless case). The only peak in this invariant mass distribution is due to kinematic effects (it peaks at about $2\times p_{T \gamma}^{\rm hard~cut} = 80$ GeV, as in the massless case). This kinematic effect is explained in detail in ref.~\cite{Catani:2018krb} for the massless result. In the bottom panel of Fig.~\ref{fig:RR} we show the relative size of the different channels (the $q\bar{q}$ and $gg$ channels) with respect to the total. For large values of $\Mgg$ the luminosity of the gluon decreases and the total contribution is (mainly) due to the $q\bar{q}$ channel. The vanishing luminosity (at large invariant masses) of the gluon explains why the $q\bar{q}$ channel dominates at large values of $\Mgg$ in the total NNLO invariant mass distribution. The prevalence of the $q\bar{q}$ channel in the tail, explains why the ratio in Fig.~\ref{fig:invmassnnlo} decreases at large values of $\Mgg$, even though the fully massive box contribution ($gg$ channel) is almost twice the massless result in the tail 
(see Fig.~\ref{fig:H2andBox} bottom panel). In absence of the two-loop massive correction, the ratio in Fig.~\ref{fig:invmassnnlo} would asymptotically approach its limit (at large invariant masses) from above, since the partonic channels containing at least one gluon vanish and the massive real corrections are almost negligible in that region. In the full result, the negative corrections coming from the $q\bar{q}$ channel (massive two-loop contribution) are still present at large invariant masses, and the ratio turns out to be negative in this kinematic region (see the ratio in Fig.~\ref{fig:H2andBox} around $\Mgg \sim 2$ TeV).

In the following, we discuss the real-virtual contribution of the one-loop NNLO massive corrections to diphoton production ($pp \rightarrow \gamma\gamma j$). In Fig.~\ref{fig:RV} we compare the invariant mass distribution of the different channels with respect to the total correction. The $q\bar{q}$ and $qg$ channels show very different behaviour (being the $qg$ initiated sub process the channel that dominates the contribution around the top quark threshold). The positive peak behaviour around the top quark threshold is also found in the box contribution when only a massive top quark is circulating in the loop \cite{Maltoni:2018zvp}. As the remaining five light flavours are also included in the loop, the destructive interference between these two types of terms dominates the box contribution, producing the typical negative peak (as it is shown in Fig.~\ref{fig:H2andBox} bottom panel). 

Here (in the real-virtual case), since the one-loop scattering amplitudes (see Fig. \ref{fig:contrib} \texttt{c)}) are interfered with the corresponding tree-level matrix elements, there is no such mixing between massive and massless quarks circulating through the loop. For large values of the invariant mass ($\Mgg > 500$ GeV) the contribution of the real-virtual term is negative, slightly enhancing the decreasing behaviour in the tail of the ratio in Fig.~\ref{fig:invmassnnlo}.
 
In Fig.~\ref{fig:RV} we also show the contribution of the whole $pp \rightarrow \gamma\gamma t \bar{t}$ sub-process. The size of the real-virtual and the double-real contributions are roughly of the same order, and they are subdominant with respect to the one-loop box and two-loop form factors.

\begin{figure}
\centering
  \includegraphics[width=0.47\textwidth]{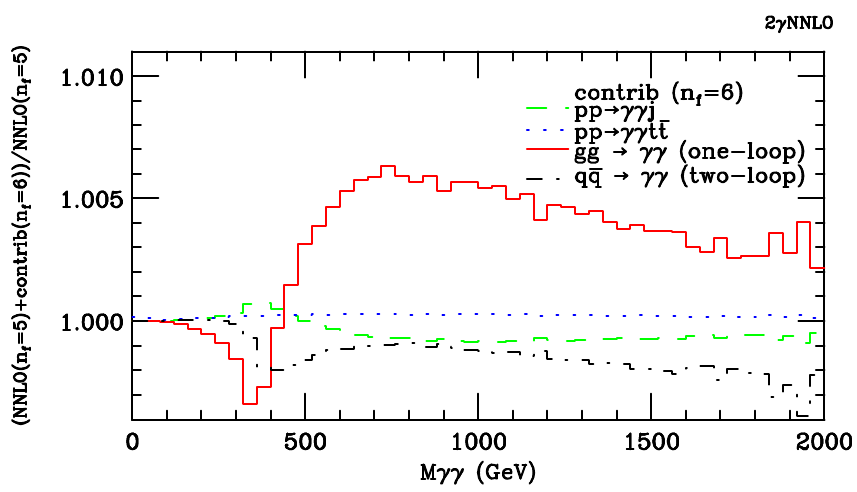}
\caption
{ \label{fig:allratios}
  { \em Ratios of each one of the massive contributions with respect to the NNLO massless cross section as a function of the invariant mass.}}
\end{figure}

Finally, in Fig.~\ref{fig:allratios} we show the ratios of each of the massive contributions (schematically drawn in Fig.~\ref{fig:contrib}) with respect to the massless NNLO differential prediction. As expected from the previous paragraphs, the two-loop $q\bar{q}$ (dot-dashed black line) and the one-loop box corrections (solid red line) dominate throughout the invariant mass range. The effect of the real-virtual contribution (dashed green line) is subdominant and reduces the size of the negative peak at the top quark threshold. The effect of the massive double real corrections (emission of two on-shell top quarks) is tiny and not relevant for the phenomenology (dotted blue line). The correction introduced by the two-loop massive contribution at large values of the invariant mass is negative and reduces the cross section. For $M_{\gamma\gamma}>$ 2 TeV it is the dominant massive effect.

\section{Conclusions}
In this paper we presented for the first time the complete NNLO QCD diphoton production taking into account the full top quark mass dependence. We presented a detailed study of the impact of the massive corrections in the invariant mass distribution around the top quark threshold. We have shown the different components of the total NNLO QCD massive result. The two most significant contributions are the one-loop ($gg\rightarrow \gamma\gamma$) box term \cite{Campbell:2016yrh} and the recently calculated two-loop ($q\bar{q}\rightarrow \gamma\gamma$) massive form factor \cite{Becchetti:2023wev}. The negative peak around the top quark threshold is, therefore, the result of (mainly) these two contributions showing the same (negative peak) behaviour. The moderated size of the positive peak introduced by the real-virtual contribution (see Fig. \ref{fig:RV}) only slightly modifies the size of the negative peak. The position of the positive peak in the ratio of Fig. \ref{fig:invmassnnlo} is the result of the two competing opposite behaviours of the two dominant contributions (see Fig. \ref{fig:H2andBox}).

The precedent discussion suggests that the massive corrections presented in this paper are relevant not only for the invariant mass region around the top mass threshold but also for larger values ($\Mgg > 2 m_t$). This kinematic region ($\Mgg \geq 500$ GeV) is of interest for BSM searches.

Recent studies of the partial N$^3$LO massive QCD results \cite{Maltoni:2018zvp,Chen:2019fla} indicate that the NLO massive corrections to the box contribution are sizable and could be relevant to the position (and size) of the positive and negative peaks in Fig. \ref{fig:invmassnnlo}. As far as subdominant contributions are concerned, the inclusion of the modulus squared of the scattering amplitudes shown in Fig. \ref{fig:contrib} \texttt{c)} (formally of $\mathcal{O}(\as^3)$) could partially reverse the effect introduced by the scattering amplitudes shown in Fig. \ref{fig:contrib} \texttt{c)} interfered with the corresponding tree-level amplitudes (as shown in Fig. \ref{fig:RV}). This is because the modulus squared of the scattering amplitudes shown in Fig. \ref{fig:contrib} \texttt{c)} will contain massive and massless flavours circulating in the loop. It is, therefore, expected that these two terms will interfere destructively (as in the case of the box), producing a negative peak behaviour around the top quark threshold. We have left the inclusion of these partial N$^3$LO massive effects to further studies.

\section*{Acknowledgements}
We would like to thank Stefano Camarda and Stefano Catani for useful comments on the manuscript.
This work is supported by the Spanish Government (Agencia Estatal de Investigaci\'on MCIN/AEI/ 10.13039/501100011033) Grant No. PID2020-114473GB-I00, and Generalitat Valenciana Grants No. PROMETEO/2021/071 and ASFAE/2022/009 (Planes Complementarios de I+D+i, Next Generation EU). 
M.B. acknowledges the financial support from the European Union Horizon 2020 research and innovation programme: High precision multi-jet dynamics at the LHC (grant agreement no. 772009). L.C. and F.C. are supported by Generalitat Valenciana GenT Excellence Programme (CIDE\-GENT/2020/011) and ILINK22045.

\bibliographystyle{elsarticle-num}
\bibliography{2fot}

\end{document}